\documentclass[letterpaper,12pt]{article}
\usepackage{color,amssymb,enumerate,float,amsmath}

\usepackage{ifpdf}
\ifpdf
	\usepackage[pdftex]{graphicx}
	\usepackage[pdftex,unicode,implicit]{hyperref}

	\hypersetup{
  	pdftitle     = {}, 
  	pdfkeywords  = {},
  	pdfauthor    = {},
  	pdfcreator   = {pdf\LaTeXe\ with package \flqq hyperref\frqq},
  	pdfproducer  = {pdf\LaTeXe\ with package \flqq hyperref\frqq},
  	pdfpagemode  = UseNone,  
  	pdffitwindow = true,  
  	unicode      = true,
  	plainpages   = true,
  	colorlinks   = true,  
  	citecolor    = black,  
  	urlcolor     = blue, 
  	linkcolor    = black
	}

\else

  \usepackage[dvips]{graphicx}

\fi 

\makeatletter
\@addtoreset{equation}{section}
\makeatother

\begin{document}
	
\thispagestyle{empty}

\begin{center}
{\bf \LARGE BFV quantization of the nonprojectable $2+1$ Ho\v{r}ava theory}
\vspace*{15mm}

{\large Jorge Bellor\'{\i}n}$^{1}$
{\large and Byron Droguett}$^{2}$
\vspace{3ex}

{\it Department of Physics, Universidad de Antofagasta, 1240000 Antofagasta, Chile.}
\vspace{3ex}

$^1${\tt jbellori@gmail.com,} \hspace{1em}
$^2${\tt byron.droguett@ua.cl}

\vspace*{15mm}
{\bf Abstract}
\begin{quotation}{\small\noindent
We show that the BFV quantization scheme can be implemented in the nonprojectable 2+1 Ho\v{r}ava theory. This opens the possibility of imposing more general gauge conditions in the quantization of this theory. The BFV quantization is based on the canonical formalism, which is suitable to incorporate the measure associated to the second-class constraints that the theory has. Special features of the Hamiltonian density and the matrix of second-class constraints allow that the system be involutive in terms of Dirac brackets, which is a nontrivial requisite for implementing the BFV formalism. We present the BRST symmetry transformations in the canonical variables. The theory is of rank one, in the classification introduced by Fradkin and Fradkina. The originally called relativistic gauge-fixing conditions of the BFV formalism can be implemented in the nonprojectable Ho\v{r}ava theory, extended to nonrelativistic forms. We show that the nonlocal gauge condition introduced in the projectable theory can be included among these gauges.
}
\end{quotation}
\end{center}


\newpage
\section{Introduction}
The Ho\v{r}ava theory of quantum gravity \cite{Horava:2009uw} can be formulated in two different versions. One is the projectable case, defined by the condition of the lapse function is a function only of time. The other version, the nonprojectable theory, has a lapse function that depends in general in time and space. The Lagrangian of the nonprojectable case was extended in Ref.~\cite{Blas:2009qj}.

An essential goal in the study of the Ho\v{r}ava theory is to prove its renormalization, which for the projectable case has been already proved \cite{Barvinsky:2015kil}. A characteristic of the proof in the projectable case is the inclusion of a nonlocal gauge-fixing condition. The nonlocal gauge leads to regular propagators for all field variables, and consequently the renormalization can be achieved by following criteria similar to the Lorentz-violating (nongravitational) field theories \cite{Anselmi:2007ri,Anselmi:2008bq,Anselmi:2008bs}. Moreover, the projectable $2+1$ Ho\v{r}ava theory has been proven to be asymptotically free \cite{Barvinsky:2017kob}. Besides these analyses, the renormalizability and other quantum aspects of the Ho\v{r}ava theory have been studied by various authors. Among them, studies on several aspects of the renormalization flow of the projectable theory, using the functional renormalization group, can be found in Refs.~\cite{Contillo:2013fua,D'Odorico:2014iha}. Even the nonprojectable case was considered in Ref.~ \cite{D'Odorico:2015yaa}. Early analysis on renormalizability are Refs.~\cite{Orlando:2009en,Benedetti:2013pya}, and further advances in the renormalization of the projectable theory has been developed in Refs.~\cite{Griffin:2017wvh,Barvinsky:2019rwn}. 

On the other hand, the proof of the renormalization of the nonprojectable case is pending. A central issue in this case is that the Hamiltonian constraint, together with an associated constraint, is a second-class constraint. This was emphasized in the nonprojectable theory, considering the extended Lagrangian of \cite{Blas:2009qj}, in the Hamiltonian analysis of Refs.~\cite{Kluson:2010nf,Donnelly:2011df,Bellorin:2011ff} (in the kinetic-conformal or critical formulation more second-class constraints arise \cite{Bellorin:2013zbp}). The presence of second-class constraints suggests to consider different schemes of quantization, since these constraints are not associated to gauge symmetries. The canonical formalism is particularly suitable for the quantization of this kind of theories. In the path integral, the measure associated to the second-class constraints is defined in terms of its matrix of Poisson brackets \cite{Fradkin1973,Senjanovic:1976br} (a symplectic-geometry approach for the measure is given in Ref.~\cite{Henneaux:1992ig}). This is also the case of the operatorial quantization based on the Dirac brackets. We presented advances in the path integral quantization of the nonprojectable $2+1$ Ho\v{r}ava theory in the canonical formalism in Ref.~\cite{Bellorin:2019gsc}.

The nonlocal gauge used in the projectable theory \cite{Barvinsky:2015kil} is a noncanonical gauge, since it involves a Lagrange multiplier. If one wants to apply a similar procedure in the nonprojectable case, then a more general quantization scheme is required, specially for being able to use noncanonical gauges. In this sense, a very general scheme of quantization of field theories in the canonical language is the so-called Batalin-Fradkin-Vilkovisky (BFV) formalism. This formalism was developed in a series of papers, Refs.~\cite{Fradkin:1975cq,Fradkin:1975sj,Fradkin:1977hw,Batalin:1977pb,Fradkin:1977xi}. In particular, in Ref.~\cite{Fradkin:1977xi} Fradkin and Fradkina extended the formalism to theories with second-class constraints and of general rank, the rank being the order of the Becchi-Rouet-Stora-Tyutin (BRST) symmetry generator and the Hamiltonian in one of the fields used in the extension of the phase space. The extension to the case with second-class constraints incorporates Dirac brackets in the path integral. The BFV formalism was originally introduced with the aim of implementing relativistic gauge-fixing conditions, like the Lorentz gauge, in the canonical quantization of relativistic theories like Yang-Mills and general relativity. The relativistic gauge-fixing conditions are noncanonical gauges (in the context of the primary canonical theory). An important achievement is that with the BFV quantization the unitarity of the S matrix can be proved, which results from showing the independence of the path integral on the gauge condition chosen and its equivalence with the canonical physical system, that is, the system without redundant or unphysical canonical degrees of freedom.

In this study we show that the BFV scheme of quantization can be implemented for the nonprojectable Ho\v{r}ava theory, incorporating its second-class constraints. To this end we follow the program of Ref.~\cite{Fradkin:1977xi}. We present explicit formulas for the $2+1$ theory. We will see that the formalism is general enough, such that it allows to impose noncanonical gauge-fixing conditions, as the kind of the nonlocal gauge condition used in the renormalization of the projectable theory, despite the fact that the BFV quantization was intented for relativistic theories with relativistic gauges. In the next section we present the BFV quantization in the general theory, and in Section 3 we apply the formalism to the linearized theory with the specific nonlocal gauge.

\section{General formalism}
Once a foliation of spatial slides along an identified time direction is given, the Ho\v{r}ava theory is formulated in terms of the Arnowitt-Deser-Misner variables $g_{ij}$, $N$ and $N_i$ over the foliation. In the nonprojectable case, which is the one we consider, $N$ is a function of time and space. The symmetry group characteristic of the theory is given by the foliation-preserving diffeomorphisms (FDiff). Given a local coordinate system, the infinitesimal form of the FDiff is
\begin{equation}
\delta t=f(t),\qquad \delta x^{i}=\zeta^{i}(t,\vec{x}).
\end{equation}
The action of the FDiff on the field variables has the form
\begin{eqnarray}
\delta N &=& 
\zeta^{k} \partial_{k} N + f \dot{N} + \dot{f}N \,, 
\label{deltaN}
\\ 
\delta N_{i} &=& 
\zeta^{k} \partial_{k} N_{i} + N_{k} \partial_{i} \zeta^{k} + \dot{\zeta}^{j} g_{ij} + f \dot{N}_{i} + \dot{f}N_{i} \,,
\\
\delta g_{ij} &=& \zeta^{k} \partial_{k} g_{ij} + 2g_{k(i}\partial_{j)} \zeta^{k} + f \dot{g}_{ij} \,.
\end{eqnarray}
Since $f(t)$ is restricted to be a function only of time, only the sector given by the general spatial diffeomorphisms ($f=0$, $\zeta^i(t,\vec{x})$ arbitrary) is a gauge symmetry in the strict sense. Hence, this symmetry must be fixed in the process of quantization, and the resulting BRST symmetry emerges from it. The Lagrangian of the nonprojectable Ho\v rava theory is given by
\begin{equation}
\mathcal{L}=
\sqrt{g} N \left( K^{ij}K_{ij}-\lambda K^{2}-\mathcal{V} \right) \,,
\label{lagrangian}
\end{equation}
where the extrinsic curvature tensor is 
\begin{equation}
K_{ij}=
\frac{1}{2N} \left(\dot{g}_{ij}-2\nabla_{(i}N_{j)}\right) \,,
\quad
K \equiv g^{ij} K_{ij} \,.
\end{equation}
The complete potential in two spatial dimensions is \cite{Sotiriou:2011dr}
\begin{eqnarray}
\mathcal{V} &=&
-\beta R-\alpha a^{2}+\alpha_{1}R^{2}+\alpha_{2}a^{4}+\alpha_{3}R a^{2}+\alpha_{4}a^{2}\nabla_{k}a^{k}
\nonumber \\ &&
+\alpha_{5} R\nabla_{k}a^{k} 
+\alpha_{6}\nabla^{l}a^{k}\nabla_{l}a_{k}+ \alpha_{7}(\nabla_{k}a^{k})^{2} \,,
\label{potencial}
\end{eqnarray}
where 
\begin{equation}
 a_i = \frac{\partial_i N}{N} 
\end{equation}
transforms as a vector under FDiff. $\lambda$, $\beta$, $\alpha$ and $\alpha_1 ,\ldots, \alpha_7$ are coupling constants. This potential has the minimal order in spatial derivatives required by power-counting renormalization, $z=2$ \cite{Horava:2009uw}, and includes the lower order terms dominant at large distances. To cast the theory in canonical formalism, the canonical variables are the pairs $(g_{ij},\pi^{ij})$, $(N,P_{N})$, where the condition $P_N = 0$ is a constraint since there are no time derivatives of $N$ in the Lagrangian (\ref{lagrangian}). Throughout this analysis we consider the asymptotic conditions $g_{ij} - \delta_{ij} = \mathcal{O}(r^{-1})$, $\pi^{ij} = \mathcal{O}(r^{-2})$ and $N - 1 = \mathcal{O}(r^{-1})$. The primary Hamiltonian, obtained by a Legendre transformation on (\ref{lagrangian}), is
\begin{equation}
H_0 =
\int d^{2}x
\sqrt{g} N \left( \frac{\pi^{ij}\pi_{ij}}{g} 
  + \frac{\lambda}{1-2\lambda} \frac{\pi^{2}}{g} + \mathcal{V} \right) \,.
\label{H0}
\end{equation}
The constraint that is going to be part of the involutive functions in the BFV quantization is given by 
\begin{equation}
\mathcal{H}_i =
- 2 g_{ij} \nabla_k \pi^{kj} = 0 \,.
\label{momentumconts}
\end{equation}
The second-class constraint of the theory are 
\begin{eqnarray}
\theta_{1} &\equiv& P_{N}=0 \,,
\label{theta1}\\
\theta_{2} &\equiv&  
\sqrt{g} N \left( \frac{\pi^{ij}\pi_{ij}}{g} 
+ \frac{\lambda}{1-2\lambda} \frac{\pi^{2}}{g} + \mathcal{V} \right) -\sqrt{g} B \,,
\label{theta2}
\end{eqnarray}
where $B$ stands for a set of total derivatives, namely
\begin{equation}
\begin{array}{rcl}
B &\equiv&
{\displaystyle -2\alpha\nabla_{k}(Na^{k})+4\alpha_{2}\nabla_{k}(Na^{2}a^{k})
+2\alpha_{3}\nabla_{k}(NRa^{k})-\alpha_{4}(\nabla^{2}(Na^{2})
}
\\[1ex] &&
{\displaystyle
-2\nabla_{l}(\nabla_{k} a^{k} N a^{l}))
-\alpha_{5}\nabla^{2}(NR)-2\alpha_{6}\nabla^{k}\nabla^{l}(N\nabla_{l}a_{k})-2\alpha_{7}\nabla^{2}(N\nabla_{l}a^{l}) \,.}
\end{array}
\label{BN}
\end{equation}

A nontrivial requisite for the BFV quantization of a given system is that it must be involutive, in this case considering the presence of second-class constraints. This means that there is a Hamiltonian density $\mathcal{H}_0$ and a set of functions $G_a$, among which the first-class constraints are included, that satisfy the algebra \cite{Fradkin:1977xi}
\begin{equation}
\{ G_a \,, G_b \}_{\mathrm{D}} = U_{ab}^c G_c \,, \quad
\{ \mathcal{H}_0 \,, G_a \}_{\mathrm{D}} = V_a^b G_b \,.
\label{definvolution}
\end{equation}
The above brackets $\{ \,, \}_{\mathrm{D}}$ are Dirac brackets, which are required by the presence of second-class constraints. They are defined in the standard way,
\begin{equation}
\left\{ F\, , R \right\}_{\mathrm{D}} =
\left\{ F\, , R \right\} 
- \left\{ F \,, \theta_A \right\} \mathbb{M}^{-1}_{AB} 
\left\{ \theta_B \,, R \right\} \,,
\quad
\mathbb{M}_{AB} = \{ \theta_A \,, \theta_B \} \,.
\end{equation}
It turns out that the canonical formulation of the nonprojectable Ho\v{r}ava theory can be given in a involutive form. To this end, we first rewrite the Hamiltonian (\ref{H0}) in a equivalent form: as a consequence of the asymptotic conditions, the spatial integral of all total derivatives in the $\theta_2$ constraint vanishes, $\int d^{2}x \sqrt{g} B = 0$. This can be used to write the Hamiltonian (\ref{H0}) as the integral of a second-class constraint,
\begin{equation}
H_{0} = \int d^{2}x \, \theta_{2} \,.
\label{primaryhamiltonian}
\end{equation}
This is key to get the involutive form. We may see how this works before making the BFV extension of the phase space: since $\theta_1$ commutes with itself at all points, the matrix of Poisson brackets between the second-class constraints acquires a triangular form,
\begin{equation}
\mathbb{M} =
\left( \begin{array}{cc}
0 & \left\{ \theta_1 , \theta_2 \right\} \\
- \left\{ \theta_1 , \theta_2 \right\} & \left\{ \theta_2 , \theta_2 \right\}
\end{array} \right) \,,
\label{secondclassmatrix}
\end{equation}
and, consequenlty, its inverse also acquires a triangular form, which we may present symbolically as
\begin{equation}
\mathbb{M}^{-1} = 
( \det\mathbb{M} )^{-1} 
\left( \begin{array}{cc}
	\mathbb{M}_{22} & - \mathbb{M}_{12} \\
	\mathbb{M}_{12} & 0
\end{array} \right)  \,.
\end{equation}
Next, since $\left\{ \mathcal{H}_i \,, \theta_1 \right\} = 0$ strongly and $\mathbb{M}^{-1}$ is triangular, the $\mathcal{H}_i$ constraint (\ref{momentumconts}) becomes involutive under the Dirac brackets. Finally, since the primary Hamiltonian density $\mathcal{H}_0$ is equivalent to a second-class constraint, automatically its Dirac bracket with any quantity is zero strongly. Thus, we obtain
\begin{eqnarray}
&&
\left\{ \mathcal{H}_i \,, \mathcal{H}_j \right\}_{\mathrm{D}} =
\left\{ \mathcal{H}_i \,, \mathcal{H}_j \right\} 
= U_{ij}^k \mathcal{H}_k \,,
\label{HiHi}
\\ &&
\left\{ \mathcal{H}_i , \mathcal{H}_0 \right\}_{\mathrm{D}} = 0 \,. 
\label{HiH0}
\end{eqnarray}
The algebra of $\mathcal{H}_i$ corresponds to the algebra of spatial diffeomorphisms at equal times,
\begin{equation}
\left\{ \mathcal{H}_i(x) , \mathcal{H}_j(y) \right\} =
  \mathcal{H}_i \frac{\partial}{\partial x^j} \delta(x-y)
+ \mathcal{H}_j \frac{\partial}{\partial x^i} \delta(x-y) \,. 
\label{spatialdiffalg}
\end{equation}

Now we proceed to perform the BFV extension of the phase space. The shift vector $N_i$ is the Lagrange multiplier of the constraint (\ref{momentumconts}). To fix the gauge symmetry of the spatial diffeomorphisms, a gauge-fixing condition, denoted by $\Phi^i = 0$, must be provided, $\pi_i$ denoting its associated Lagrange multiplier. The canonical pair $(N^i , \pi_i)$ is incorporated to the phase space. Due to the Dirac brackets (\ref{HiHi}) and (\ref{HiH0}), and from the fact that none of the objects $\mathcal{H}_i$, $\theta_A$, $\mathcal{H}_0$ depends on the pair $(N^i,\pi_i)$, the functions $G_a = ( \mathcal{H}_i , \pi_i )$ and $\mathcal{H}_0$ are involutive: they satisfy (\ref{definvolution}), where the nonzero components $U_{ab}^c$ are read from (\ref{spatialdiffalg}), and $V_a^b = 0$.

Next, for each function $G_a$ we add the canonical pair of BFV ghosts $(\eta^a , \mathcal{P}_a)$, which are Grassmann variables. It is sometime convenient to split the ghosts in the way $\eta^a = (\eta_1^i , \eta_2^i )$, $\mathcal{P}_a = (\mathcal{P}^1_i , \mathcal{P}^2_i )$, hence we use indices like $a,b,c\ldots$ for the unsplitted variables. Thus, the full phase space is  given by the pairs $(g_{ij},\pi^{ij})$, $(N, P_{N})$, $(N^{i},\pi_{i})$, $(\eta^{a},\mathcal{P}_{a})$. The definition of the Poisson brackets in the extended phase space is ($q^A,\pi_A$ stands for all the canonical pairs)
\begin{equation}
 \{ F \,, R \} = 
 \frac{\delta_r F}{\delta q^A} \frac{\delta_l R}{\delta \pi_A}
 -
 (-1)^{n_R n_F}
 \frac{\delta_r R}{\delta q^A} \frac{\delta_l F}{\delta \pi_A} \,,  
 \label{poisson}
\end{equation} 
where $r,l$ denote the right and left derivatives and $n_R = 0$ ($=1$) if $R$ is a boson (fermion).

The BFV path integral of the nonprojectable Ho\v{r}ava theory is given by
\begin{equation}
Z_{\Psi}=
\int \mathcal{D}V  \delta(\theta_{1}) \delta(\theta_{2}) e^{iS},
\label{Z}
\end{equation}
where the measure and the action are given, respectively, by
\begin{eqnarray}
&&
\mathcal{D}V = 
\mathcal{D} g_{ij} \mathcal{D}\pi^{ij} \mathcal{D}N \mathcal{D}P_{N}\mathcal{D} N^{k}\mathcal{D}\pi_{k}\mathcal{D}\eta^{a}\mathcal{D}\mathcal{P}_{a} \times  \sqrt{\det \mathbb{M}} \,,
\label{medida}
\\ &&
S=
\int dt d^{2}x \left( P_N \dot{N} + \pi^{ij} \dot{g_{ij}} 
+ \pi_{k} \dot{N}^{k} + \mathcal{P}_{a}\dot{\eta}^{a} - \mathcal{H}_{\Psi}
\right) \,.
\label{scan} 
\end{eqnarray}
The factor $\sqrt{\det \mathbb{M}}$ os the part of the measure associated to the second-class constraints \cite{Fradkin1973,Senjanovic:1976br}. The quantum gauge-fixed Hamiltonian density is
\begin{equation}
\mathcal{H}_{\Psi}=\mathcal{H}_{1}+\{\Psi,\Omega\}_{\mathrm{D}} \,,
\end{equation}
where $\Omega$ is a generator of a kind of BRST symmetry and $\Psi$ is a gauge-fixing fermionic function, which can have a general dependence on the extended phase space, $\Psi = \Psi(g_{ij},\pi^{ij},N,N^i,$ $\pi_i,\eta^a,\mathcal{P}_a)$. The objects that arise in the gauge-fixed Hamiltonian are defined by (for simplicity of the notation, we write densities)
\begin{eqnarray}
&&
\Omega = 
G_{a}\eta^{a} 
+ \sum_{k=1}^{s} \mathcal{P}_{b_k} \cdots \mathcal{P}_{b_1} 
  \Omega^{b_1 \cdots b_k},
\label{omega}
\\ &&
\mathcal{H}_{1} = \mathcal{H}_{0} 
+ \sum_{k=1}^{s} \mathcal{P}_{b_k} \cdots \mathcal{P}_{b_1}
\mathcal{H}_1^{b_1\cdots b_k} \,.
\label{h1}
\end{eqnarray}
Fradkin and Fradkina \cite{Fradkin:1977xi} defined a theory to be of rank $s$ if it is possible to define the expansions of $\Omega$ and $\mathcal{H}_1$ as polynomials of degree $s$ in the $\mathcal{P}$'s, that is, if $\Omega^{b_1\cdots b_k} = 0$ for $k > s$, and similarly for $\mathcal{H}_1$, as shown in (\ref{omega}) -- (\ref{h1}). The coefficient functions of the first order in $\mathcal{P}_a$ are given by
\begin{equation}
 \Omega^a = - \frac{1}{2} U_{bc}^a \eta^b \eta^c  \,,
 \quad
 \mathcal{H}_1^a = V^a_b \eta^b \,,
 \label{firstorder}
\end{equation}
and, starting from them, the higher order coefficients functions $\Omega^{a_1 \cdots a_k}$ and $\mathcal{H}_1^{a_1 \cdots a_k }$, up to the rank $s$, are given by recurrence relations shown in Ref.~\cite{Fradkin:1977xi}. The final generator $\Omega$ and the gauge-fixed Hamiltonian $\mathcal{H}_{\Psi}$ must satisfy the following conditions,
\begin{eqnarray}
&&
\{\Omega \,,\Omega\}_{\mathrm{D}} =0 \,,
\label{omeganilpotent}
\\ &&
\{\mathcal{H}_{1} \,,\Omega\}_{\mathrm{D}} =0 \,.
\label{omegah1**}
\end{eqnarray}
Since $\Omega$ is a fermion, Eq.~(\ref{omeganilpotent}) is a nontrivial condition, according to the definition (\ref{poisson}).

We obtain that the nonprojectable Ho\v{r}ava theory can be casted as a theory of range one, such that the coefficient functions of the generator $\Omega$ and the Hamiltonian are given by (\ref{firstorder}). Let us set
\begin{eqnarray}
&&
\Omega = 
G_{a} \eta^{a} - \frac{1}{2} U^{c}_{ab} \eta^{a} \eta^{b}  \mathcal{P}_{c}
= 
\mathcal{H}_{k} \eta_{1}^{k} + \pi_{k} \eta_{2}^{k} 
- \frac{1}{2} U^{k}_{ij} \eta_{1}^{i} \eta_{1}^{j} \mathcal{P}_{k}^{1} \,,
\\ &&
\mathcal{H}_{1} = \mathcal{H}_{0} + V_b^a \mathcal{P}_a \eta^b\,.
\end{eqnarray} 
The Dirac brackets of these objects are
\begin{eqnarray}
\{\Omega \,, \Omega\}_{\mathrm{D}} &=& 
\{ \mathcal{H}_i \eta^i_1 \,, \mathcal{H}_j \eta^j_1 \}_{\mathrm{D}} 
- \{ \mathcal{H}_i \eta^i_1 \,,  
 U^{k}_{mn} \eta_{1}^{m} \eta_{1}^{n} \mathcal{P}_{k}^{1} \}_{\mathrm{D}} 
+ \frac{1}{4} \{ U^{k}_{ij} \eta_{1}^{i} \eta_{1}^{j} \mathcal{P}_{k}^{1} \,,  
  U^{l}_{mn} \eta_{1}^{m} \eta_{1}^{n} \mathcal{P}_{l}^{1} \}_{\mathrm{D}} \,,
\nonumber \\ \label{OMEGAOMEGA}
\\
\{\mathcal{H}_{1} \,, \Omega\}_{\mathrm{D}} &=& 
\{ \mathcal{H}_{0} \,, \Omega \}_{\mathrm{D}} 
+ \{ V^{b}_{a} \mathcal{P}_{a} \eta^{b} \,, \Omega \}_{\mathrm{D}}
 \,.
\label{H1omega}
\end{eqnarray}
In the right-hand side of Eq.~(\ref{OMEGAOMEGA}), the first and second brackets are identical,
\begin{equation}
\{ \mathcal{H}_i \eta^i_1 \,, \mathcal{H}_j \eta^j_1 \}_{\mathrm{D}} = 
\{ \mathcal{H}_i \eta^i_1 \,,  
U^{k}_{mn} \eta_{1}^{m} \eta_{1}^{n} \mathcal{P}_{k}^{1} \}_{\mathrm{D}} = 
U^i_{jk} \eta_{1}^{j} \eta_{1}^{k} \mathcal{H}_i \,,
\end{equation}
hence they cancel, whereas the third bracket is proportional to the structure $\eta_1^j \eta_1^m \eta_1^n U^k_{ij} U^i_{mn}$, which is zero by the Jacobi identity. The Dirac bracket (\ref{H1omega}) is zero identically, since for the nonprojectable Ho\v{r}ava theory all the $V^{a}_{b}$ are zero, and the Hamiltonian density  $\mathcal{H}_{0}$ is equivalent to a second-class constraint, hence its Dirac bracket with any quantity is automatically zero. Thus, the BFV quantization of the nonprojectable Ho\v{r}ava theory is well posed, and the theory is of rank one. The BFV gauge-fixed Hamiltonian takes the form
 \begin{equation}
\mathcal{H}_{\Psi} =
\mathcal{H}_{0}
+ \left\{ \Psi \,, \mathcal{H}_{k} \eta_{1}^{k} + \pi_{k} \eta_{2}^{k} 
- \frac{1}{2} U^{k}_{ij} \eta_{1}^{i} \eta_{1}^{j} \mathcal{P}_{k}^{1} \right\}_{\mathrm{D}} \,.
\label{gaugefixing}
\end{equation}

In the original BFV formulation, a specific form of the gauge-fixing function $\Psi$ was introduced with the aim of quantizing theories using relativistic gauges. It turns out that this form is suitable for the Ho\v{r}ava theory, regardless of the fact that it is not a relativistic theory. By adpating the notation to the Ho\v{r}ava theory, the relativistic gauge considered in the BFV formulation has the general form
\begin{equation}
\Phi^i = - \dot{N}^i 
+ \chi^i(g_{ij},N,\pi^{ij}, N^i, \pi_i , \eta^a, \mathcal{P}_a ) \,,
\label{relativisticgaugephi}
\end{equation} 
where $\chi^i$ is the chosen part of the gauge fixing. Since the functional part $\chi^i$ is left free, one may use this form of gauge fixing both for relativistic and nonrelativistic gauges. Given $\chi^i$, and identifying $\chi^{a}=(N^{i},\chi^{i})$, the specific BFV fermionic gauge-fixing function is
\begin{equation}
\Psi = \mathcal{P}_{a}\chi^{a}=\mathcal{P}_{i}^{1}N^{i}+\mathcal{P}_{i}^{2}\chi^{i} \,.
\label{relativisticgaugepsi}
\end{equation}   
In this case the gauge-fixed Hamiltonian takes the form
\begin{equation}
\mathcal{H}_{\Psi} =
\mathcal{H}_{0} 
+ \mathcal{P}_{k}^{1} \eta_{2}^{k} + \mathcal{H}_{k} N^{k}
+ U^{k}_{ij} N^{i} \eta_{1}^{j} \mathcal{P}^{1}_{k}
+ \{ \mathcal{P}_{i}^{2} \chi^{i} \,, \Omega\}_{\mathrm{D}} \,.
\label{genhamiltonian}
\end{equation}
If $\chi^i$ does not depend on the BFV ghosts $\eta^a,\mathcal{P}_a$, the bracket indicated in (\ref{genhamiltonian}) simplifies to $\left\{ \mathcal{P}_{i}^{2} \chi^{i} \,, 
\mathcal{H}_{k} \eta_{1}^{k} + \pi_{k} \eta_{2}^{k} \right\}_{\mathrm{D}}$.

The BRST symmetry in the BFV formalism is implemented with the generator $\Omega$ in the form of transformations with Dirac brackets, 
\begin{equation}
 \tilde{\varphi} = \varphi + \{ \varphi \,, \Omega \}_{\mathrm{D}} \mu \,,
 \label{brstdef}
\end{equation}
where $\varphi$ represents each one of the canonical fields of the fully extended canonical phase space, and $\mu$ is the fermionic global parameter of the transformation. The nilpotency of $\Omega$ is obtained by means of the Jacobi identity: for an object $\Upsilon$,
\begin{equation}
 2 \{ \{ \Upsilon \,, \Omega \}_{\mathrm{D}} \,, \Omega \}_{\mathrm{D}} +
 \{ \{ \Omega \,, \Omega \}_{\mathrm{D}} \,, \Upsilon \}_{\mathrm{D}} = 0 \,, 
\end{equation}
the second term being zero by (\ref{omeganilpotent}).

It is direct to check the BRST symmetry of the gauge-fixed Hamiltonian of the nonprojectable Ho\v{r}ava theory (\ref{gaugefixing}). Its transformation is
\begin{equation}
 \delta_{\Omega} \mathcal{H}_{\Psi} = 
 \{ \mathcal{H}_{0} \,,\Omega \}_{\mathrm{D}}
 +\{ \{\Psi \,, \Omega \}_{\mathrm{D}} \,, \Omega\}_{\mathrm{D}} \,.
\label{transfH}
\end{equation}
The first bracket in the right-hand side is zero since $\mathcal{H}_0$ is equivalent to a second-class constraint, and the last bracket is zero by the nilpotency of $\Omega$. Explicitly, the BRST transformations (\ref{brstdef}) of the theory take the form 
\begin{equation}
\begin{array}{ll}
\delta_\Omega g_{ij} =
2g_{k(i}\nabla_{j)}\eta_1^{k}\mu \,,
\hspace*{4em} &
\delta_\Omega \pi^{ij} =
-2\pi^{k(i}\nabla_k\eta_1^{j)} \mu + \nabla_k(\pi^{ij}\eta_1^{k}) \mu \,,
\\[1ex]
{\displaystyle
\delta_\Omega N =
- \mathbb{M}^{-1}_{12} \{ \theta_2 \,, \mathcal{H}_k\eta_1^{k} \} \mu \,,
}
&
\delta_\Omega P_N = 0 \,,
\\[2ex]
\delta_\Omega N^{k} = \eta_{2}^{k} \mu \,,
&
\delta_\Omega \pi_{k} = 0 \,,
\\[1ex] {\displaystyle
\delta_\Omega\eta_{1}^{i} = 
- \frac{1}{2} U^{i}_{jk} \eta^{j}_{1} \eta^{k}_{1} \, \mu } \,,
& {\displaystyle
\delta_\Omega\mathcal{P}_{i}^{1} =
\mathcal{H}_i \mu
- U^{k}_{ij} \eta_{1}^j \mathcal{P}^{1}_{k} \mu } \,, 
\\[1ex]
\delta_\Omega \eta_{2}^{k} = 0 \,,
&
\delta_\Omega \mathcal{P}_{k}^{2} = 
\pi_{k} \mu \,.
\end{array}
\label{brst}
\end{equation}
The transformations of $g_{ij}$ and $\pi^{ij}$ are spatial diffeomorphisms with the vector argument composed of the fermions $\eta_1^k \mu$. We make a comment about the nonzero transformation of $N$: the canonical action in Eq.~(\ref{scan}) is invariant under the BRST transformations (\ref{brst}) when it is evaluated on the constrained phase space. Indeed, the kinetic term produces a transformation $\delta_\Omega (P_N \dot{N}) \sim P_N \partial_t \delta_\Omega N$, that vanishes in the constrained phase space, where $P_N = 0$. 

Finally, we comment that, according to Eq.~(\ref{secondclassmatrix}), for the Ho\v{r}ava theory the part of the measure  associated to second-class constraints takes the form
\begin{equation}
\sqrt{\det\{\theta_{p},\theta_{q}\}}=\det\{\theta_{1},\theta_{2}\},
\label{sprtbra}
\end{equation}
hence we can include the measure in the Lagrangian with a pair of fermionic ghosts $\varepsilon$ and $\bar{\varepsilon}$ \cite{Bellorin:2019gsc},
\begin{equation}
\det\{ \theta_{1} \,, \theta_{2} \}
= \int \mathcal{D}\bar{\varepsilon} \, \mathcal{D} \varepsilon
\exp\left( i \int dt d^2x\, \bar{\varepsilon}\, \{ \theta_{1} \,, \theta_{2} \}\, \varepsilon \right) \,.
\label{gostsecond}
\end{equation}

\section{Linearized theory}
We introduce perturbative quantum variables around the Minkowski spacetime, according to $g_{ij} - \delta_{ij} =  h_{ij}$, $N - 1 = n$, and the rest of canonical variables are considered of perturbative order: $\pi^{ij} = p^{ij}$, $P_N = p_n$ and $N^i = n^i$. We make the change of notation $\eta_1^i \rightarrow C^i$, $\eta_2^i \rightarrow \mathcal{P}^i$, $\mathcal{P}^1_i \rightarrow \bar{\mathcal{P}}_i$ and $\mathcal{P}^2_i \rightarrow \bar{C}_i$, such that 
$\mathcal{P}_{a}\dot{\eta}^{a} = \bar{\mathcal{P}}_{k}\dot{C}^{k}+\bar{C}_{k}\dot{\mathcal{P}}^{k}$.
The perturbative BFV path integral is given by Eq.~(\ref{Z}), where the measure and the action are given, respectively, by
\begin{eqnarray}
&&
\mathcal{D}V = 
\mathcal{D} h_{ij} \mathcal{D}p^{ij} \mathcal{D}n \mathcal{D}p_{n}\mathcal{D} n^{k}\mathcal{D}\pi_{k}\mathcal{D}\eta^{a}\mathcal{D}\mathcal{P}_{a} \times \det\{\theta_1 \,,\theta_2 \} \,,
\label{medidapertur}
\\ &&
S=
\int dt d^{2}x 
\left( p_n \dot{n} + p^{ij}\dot{h}_{ij} 
+ \pi_{k} \dot{n}^{k} + \mathcal{P}_{a} \dot{\eta}^{a} 
- \mathcal{H}_{\Psi} \right)
\label{scanpertur} \,.
\end{eqnarray}
The constraint $\mathcal{H}_i$ and the Hamiltonian density $\mathcal{H}_0$, taken from the primary Hamiltonian (\ref{H0}), at quadratic order become
\begin{eqnarray}
\mathcal{H}_{j} &=&
-2\delta_{ij}\partial_{k}p^{ki}-2\partial_{k}(h_{ij}p^{ki})+p^{kl}\partial_{j}h_{kl} \,,
\\
\mathcal{H}_{0} &=&
p^{ij}p^{ij}+\frac{\lambda}{1-2\lambda}p^{2}
+ \beta \left( \frac{1}{4}h\partial^{2}h 
- \frac{1}{4}h_{ij}\partial^{2}h_{ij} 
+ \frac{1}{2}h_{ij}\partial_{i}\partial_{k}h_{jk}
+ n\partial^{2}h 
- n\partial_{i}\partial_{j}h_{ij} \right.
\nonumber \\ && \left.
- \frac{1}{2}h\partial_{i}\partial_{j}h_{ij} \right)
+\alpha n \partial^{2}n 
+ \alpha_{1} \left( h\partial^{4}h
+h_{kl}\partial_{k}\partial_{l}\partial_{i}\partial_{j}h_{ij}
-2h\partial^{2}\partial_{i}\partial_{j}h_{ij} \right)
\nonumber \\ &&
-\alpha_{5} \left( n\partial^{4}h
- n\partial^{2}\partial_{i}\partial_{j}h_{ij} \right)
+(\alpha_{6}+\alpha_{7})n\partial^{4}n \,.
\end{eqnarray}

The sector of the BFV ghosts that play the role of canonical momenta can be integrated, such that the canonical formulation of this sector can be translated to a Lagrangian formulation, at least for the case of the gauge fixing condition (\ref{relativisticgaugephi}) -- (\ref{relativisticgaugepsi}), and if the sector $\chi^i$ of the gauge fixing depends only on the primary canonical variables $h_{ij},p^{ij},n$. Under these conditions, we examine the dependence of the Hamiltonian (\ref{genhamiltonian}) on the ghost canonical momenta $\mathcal{P}^i, \bar{\mathcal{P}}_i$. The Dirac bracket indicated in Eq.~(\ref{genhamiltonian}) reduces to
\begin{equation}
  \left\{ \bar{C}_{i} \chi^{i} \,, 
  \mathcal{H}_{k} C^{k} + \pi_{k} \mathcal{P}^{k} \right\}_{\mathrm{D}} 
  =
  \left\{ \bar{C}_{i} \chi^{i} \,, \mathcal{H}_{k} C^{k} \right\}_{\mathrm{D}}
  + \chi^i \pi_i \,,
\end{equation}
such that it does not depend on $\mathcal{P}^i,\bar{\mathcal{P}}_i$. Next, the second term in Eq.~(\ref{genhamiltonian}) is $\bar{\mathcal{P}}_i \mathcal{P}^i$, whereas the fourth term is of cubic order in perturbations. After these considerations, the terms of the quadratic action that depend on $\mathcal{P}^i , \bar{\mathcal{P}}_i$ are
\begin{equation}
\bar{\mathcal{P}}_{k}\dot{C}^{k} + \bar{C}_{k}\dot{\mathcal{P}}^{k} - \bar{\mathcal{P}}_{k}\mathcal{P}^{k} 
=
\left( \bar{ \mathcal{P}}_{k} + \dot{\bar{{C}}}_{k} \right) 
\left( -\mathcal{P}^{k} + \dot{C}^{k} \right)
+ \dot{C}^{k} \dot{\bar{C}}_{k}
 \,.
\end{equation}
The variables between brackets can be integrated out without consequences in the path integral. The second-class constraint $\theta_2$ can be incorporated to the Lagrangian by means of a Lagrange multiplier, which we denote by $a$. Therefore, the action up to second order in perturbations, using Eq.~(\ref{gostsecond}), is given by
\begin{eqnarray}
S &=&
\int dt d^{2}x \left( 
p^{ij}\dot{h}_{ij} + \pi_{k}\dot{n}^{k} + \dot{C}^{k} \dot{\bar{C}}_{k}
- \mathcal{H}_{0} - \mathcal{H}_{k}n^{k} 
\right.
\nonumber\\ && \left.
- \left\{ \bar{C}_{i} \chi^{i} \,, 
  \mathcal{H}_{k} C^{k} \right\}_{\mathrm{D}} - \chi^i \pi_i + a\theta_2 + \bar{\varepsilon} \{\theta_1 \,, \theta_2 \} \varepsilon \right) \,.
\end{eqnarray}

An example of noncanonical gauge condition that can be adapted to the BFV formalism in the form (\ref{relativisticgaugephi}) is the nonlocal gauge introduced in the projectable Ho\v{r}ava theory in Ref.~\cite{Barvinsky:2015kil}. Since the projectable theory is a theory without second-class constraints, to define the quantum theory a Faddeev-Popov procedure is implemented to fix the gauge freedom associated to spatial diffeomorphisms, which is performed in the Lagrangian version directly. The gauge-fixing term added to the Lagrangian has the form \cite{Barvinsky:2015kil}
\begin{equation}
 \mathcal{L}_{gf} \sim F^i \mathcal{O}_{ij} F^j,
 \label{faddeevpopov}
\end{equation}
where
\begin{equation}
 F^i = \dot{n}^i
 -\frac{1}{2\sigma}\partial^{2}\partial_{k}h_{ik}
 +\frac{\lambda}{2\sigma}(1+ \xi)\partial^{2}\partial_{i}h
 -\frac{\xi}{2\sigma}\partial_{i}\partial_{j}\partial_{k} h_{jk} \,,
\end{equation}
$\xi,\sigma$ are arbitrary constants. The nonlocality of the gauge is due to the operator $\mathcal{O}_{ij}$. In Ref.~\cite{Barvinsky:2015kil} it is discussed that the most general nonlocal operator that preserve the degree of space-time anisotropy of the $2+1$ theory ($z=2$) has the form
\begin{equation}
 \mathcal{O}_{ij} = 
 - \left[ \delta_{ij} \partial^2 + \xi \partial_i \partial_j \right]^{-1} \,.
 \label{nonlocal}
\end{equation}
In order to make this anisotropic (nonrelativistic) and nonlocal gauge-fixing condition compatible with the BFV gauge-fixing (\ref{relativisticgaugephi}), we must include the operator $\mathcal{O}_{ij}$ in the form of $\Phi^i$. For our purposes the explicit form of the operator $\mathcal{O}_{ij}$ is not needed. We just denote by $\mathcal{O}$ the nonlocal operator acting on $F^i$ (the square root of $\mathcal{O}_{ij}$). The important property of $\mathcal{O}$ in our computations is that it is independent of the fields, as seen in (\ref{nonlocal}). We may perform a canonical transformation in the BFV formalism, given by
\begin{equation}
 \mathcal{O} n^i \rightarrow n^i \,, 
 \quad
 \pi_i \mathcal{O}^{-1} \rightarrow \pi_i \,.
\end{equation}
Since $\mathcal{O}$ is field-independent, this transformation does not alter the measure of the path integral. Thus, the nonlocal gauge fixing condition gets the form (\ref{relativisticgaugephi}), with
\begin{eqnarray}
\chi^{i}{} &=&
\mathcal{O} \left(
-\frac{1}{2\sigma}\partial^{2}\partial_{k}h_{ik}
+\frac{\lambda}{2\sigma}(1+\xi)\partial^{2}\partial_{i}h
-\frac{\xi}{2\sigma}\partial_{i}\partial_{j}\partial_{k} h_{jk}
\right) \,.
\end{eqnarray}
The quantum BFV action with this nonlocal gauge is given by
\begin{eqnarray}
 S &=&
 \int dt d^{2}x 
 \left( p^{ij} \dot{h}_{ij} 
 + \pi_{k}\dot{n}^{k} 
 - \frac{1}{2\sigma} \pi_{k} \mathcal{O}^2 \left( \partial^{2}\partial_{l}h_{kl}
 - \lambda (1+ \xi )\partial^{2}\partial_{k}h
 + \xi \partial_{k}\partial_{j}\partial_{l} h_{jl} \right) \right.
 \nonumber \\ &&
 -\mathcal{H}_{0} 
 - \mathcal{H}_{k} \mathcal{O}^{-1} n^{k}
 -\dot{\bar{C}}_{k}\dot{C}^{k}
 -\frac{1}{\sigma} \Big( \frac{1}{2} - \lambda + \xi (1-\lambda) \Big) C^{k} \mathcal{O} \partial_{k} \partial^{2} \partial_{i} \bar{C}_{i}
 \nonumber \\ && \left.
 +\frac{1}{2\sigma}C^{k} \mathcal{O} \partial^{4}\bar{C}_{k}
  +a \theta_2
  + \bar{\varepsilon} \{ \theta_1 \,, \theta_2 \} \varepsilon \right) \,.
  \label{finalaction}
 \end{eqnarray}
The final Lagrangian is nonlocal, as the quantum Lagrangian used in Ref.~\cite{Barvinsky:2015kil} in the projectable theory.


\section{Conclusions}
We find that the BFV quantization, based on the canonical formalism, is well posed for the nonprojectable Ho\v{r}ava theory. The formalism allows the incorporation of the second-class constraints, according to the extension done in Ref.~\cite{Fradkin:1977xi}. Its application is rather nontrivial due to the neccesity of using Dirac brackets, and obtaining afterwards an involutive system in terms of these brackets. We have presented explicit formulae for the case of the $2+1$ dimensional theory. One advantage of this scheme of quantization is the possibility of using more types of gauge-fixing conditions, in particular, gauges that are noncanonical in the primary canonical formulation of the theory. In this sense, the applicability of the BFV formalism goes beyond the relativistic theories and relativistic gauges. We have shown as an example the nonlocal gauge-fixing condition introduced in the case of the projectable Ho\v{r}ava theory \cite{Barvinsky:2015kil}. Despite this, the irregularness of the propagators of the auxiliary fields associated to the second-class constraints persists. This can be seen from Eq.~(\ref{finalaction}): unlike the BFV ghosts $\bar{C}_i$, $C^i$ that get a kinetic term, no kinetic terms are generated for the fields $a,\varepsilon,\bar{\varepsilon}$, whose role is to ensure the correct implementation of the second-class constraints (we pointed out an analogue behavior in Ref.~\cite{Bellorin:2019gsc}). This connects with another feature we have found: the final BRST symmetry in the BFV formulation holds only in the constrained phase space, that is, after the second-class constraints are imposed.

\section*{Acknowledgements}
B.~D.~is partially supported by the CONICYT PFCHA/DOCTORADO BECAS  
CHILE /2019 - 21190398, and by the grant PROYECTO ANT1956 of Universidad de Antofagasta, Chile.


   \end{document}